\documentclass[preprint,showpacs,aps]{revtex4}
\usepackage{graphicx}

\def\RR{\hbox{{\rm I}\kern-.2em\hbox{\rm R}}}
\def\pRR{\hbox{{\tiny \rm I}\kern-.1em\hbox{{\tiny \rm R}}}}

\def\NN{\hbox{I\kern-.2em\hbox{N}}}

\begin{document}

\title{Igniting homogeneous nucleation}
\author{ J. C. Neu\cite{neu:email}}
\affiliation{Department of Mathematics, University of California
at Berkeley, Berkeley, CA 94720, USA}
\author{ L. L. Bonilla\cite{bonilla:email}}
\affiliation{Escuela Polit\'ecnica Superior, Universidad Carlos
III de Madrid, Avda.\ Universidad 30, E-28911 Legan{\'e}s, Spain}
\author{ A. Carpio\cite{carpio:email}}
\affiliation{Departamento de Matem\'atica Aplicada, Universidad
Complutense de Madrid, E-28040 Madrid, Spain}
\date{\today}

\begin{abstract} 
Transient homogeneous nucleation is studied in the limit of large critical sizes. Starting 
from pure monomers, three eras of 
transient nucleation are characterized in the classic Becker-D\"oring kinetic equations with
two different models of discrete diffusivity: the classic Turnbull-Fisher formula and an
expression describing thermally driven growth of the nucleus. The latter diffusivity yields
time lags for nucleation which are much closer to values measured in experiments with
disilicate glasses. After an initial stage in which the
number of monomers decreases, many clusters of small
size are produced and a continuous size distribution is created. During the second era,
nucleii are increasing steadily in size in such a way that their distribution
appears as a wave front advancing towards the critical size for steady
nucleation. The nucleation rate at critical size is negligible during this era.
After the wave front reaches critical size, it ignites the creation  of supercritical clusters 
at a rate that increases monotonically until its steady value is reached. Analytical formulas 
for the transient nucleation rate and the time lag are obtained that improve
classical ones and compare very well with direct numerical solutions.
\end{abstract}

\maketitle

\section{Introduction}
\label{sec-introduction}
Homogeneous nucleation occurs in many examples of first order phase 
transitions \cite{ll10} such as condensation of liquid droplets from a
supersaturated vapor, glass-to-crystal transformations \cite{kel83}, crystal nucleation 
in undercooled liquids \cite{kel91}, and in polymers \cite{cap03}, colloidal 
crystallization \cite{gas01}, growth of spherical aggregates beyond the critical 
micelle concentration (CMC) \cite{isr91,neu02}, and the segregation by coarsening 
of binary alloys quenched into the miscibility gap \cite{LS,XH,mar96}. In condensed 
systems, a long time elapses
before the nucleation rate (at which stable nucleii larger than the critical size are 
generated) reaches a steady state, therefore these systems offer excellent 
opportunities to study time-dependent nucleation \cite{kel91}. 

Understanding the kinetics of nucleation and growth beyond the determination 
of the steady-state nucleation rate is a task of great importance and
not yet completely accomplished. For example, it is desirable to obtain a simple asymptotic 
description of the transient until the steady-state nucleation stage sets in. Moreover, there 
is no clear distinction between nucleation and growth, and an unified theory of both processes 
does not exist \cite{kel91} despite recent attempts at bridging the gap between nucleation 
and late-stage coarsening theories \cite{pen97,juanjo,nie03}.

In this paper, we consider the problem of describing the approach to steady-state
nucleation within the classical nucleation theory \cite{kel91}. Thus our starting point
is the Becker-D\"oring (BD) discrete  kinetic model of nucleation and indefinite growth 
of a stable phase from a metastable state \cite{kel91,pen83,gan99}. The BD model contains 
two kinetic rate constants that are related to each other by assuming detailed balance. 
To complete the description of the BD equations (BDE), a model for one of the 
rate constants, usually a discrete diffusivity describing the rate at which a cluster loses 
one monomer, is needed. In the classical theory, the discrete diffusivity is given by the
Turnbull-Fisher (TF) expression which assumes that a monomer has to overcome an
activation energy barrier for its transfer across the interface of a cluster. The TF discrete
diffusivity is therefore proportional to the surface area of the cluster 
\cite{tur49}. Other models are selected so as to yield the known expression for the 
adiabatic growth of a nucleus of critical size by either diffusion or by heat transfer. The 
discrete diffusivity of these later models is proportional to the cluster radius.

No matter which discrete diffusivity is used, starting from an initial condition of pure 
monomers surpassing the CMC, we expect that cluster size increases and stable supercritical 
nucleii are formed at a {\em nucleation} rate that will eventually become stationary at an 
exponentially small value. After the stationary nucleation has set in, the supercritical clusters
continue growing, and the discrete diffusivity of the BDE can be ignored in the description
of their growth, which is a pure convection in the space of cluster size. For precipitation 
processes, this will eventually result in late stage coarsening which we will not study in the 
present paper. 

The small parameter that informs our asymptotic analysis is $\epsilon=k_{c}^{-1/3}$, 
where $k_{c}$ is the number of monomers in the critical nucleus, a `large' 
quantity that ranges between 20 and 1000 for common materials \cite{kel91}. Using other 
small parameters, such as the supersaturation, yields particular cases of 
our results (cf.\ Chapter 2 by Neu and Bonilla in Ref. \cite{cap03}). The analysis of the 
BDE in the limit as $\epsilon\to 0$ distinguishes three well defined 
stages or eras in the approach to the stationary nucleation rate. Starting from the initial 
state of pure monomers, a continuous distribution
of cluster sizes is established at the monomer's expense during the first era.
During the second era, the clusters grow to the critical size in such a way that 
their size distribution is a traveling wave front in size space. As this wave reaches
the critical size, the formation of supercritical nucleii starts, nucleation is {\em ignited}, 
and the nucleation rate increases from zero to its stationary value during the third era. 
We have obtained two different expressions for the nucleation rate (which is of paramount 
importance to compare with experiments): (I) a general expression in terms of the 
instantaneous location of the wave front and its instantaneous width, which solve two given 
differential equations, and (II) a more explicit description of the nucleation rate in terms of 
the solution of the linearized wave front position with an origin of time
at the time $t_{M}$ needed for the exact wave front to advance from pure monomers to a 
certain near critical size. Numerical solution of the model confirms all the theoretical 
predictions. 

Most previous studies of transient nucleation considered the Zeldovich-Frenkel equation 
(ZFE), which is a Fokker-Planck-type equation resulting from taking the continuum limit 
of the BDEÊ\cite{zel43}. Zeldovich \cite{zel43} set the discrete diffusivity equal to its
value at the critical cluster size and used a parabolic approximation for the variation of the 
free energy. The resulting expression for the transient nucleation rate was rather inaccurate
\cite{kel83}. Until the mid 1980s, work on the ZFE was based on similarly uncontrolled 
approximations \cite{kas69}. Some of them gave expressions for the nucleation rate and 
time lag close to the values obtained by numerically solving the BDE for particular parameter values,
but were far off for other parameter ranges \cite{wu96}. Asymptotic theories for the ZFE
were elaborated later \cite{tri87,shn87,shi90}. There are two main differences between 
asymptotic results obtained for the discrete BDE and those obtained for the continuum ZFE: 
(i) the time lags for transient nucleation are different, as explained by Wu \cite{wu96}, and 
(ii) the width of the wave front and the time to ignition are different 
(wider for the ZFE). Nevertheless, other magnitudes such as relaxation times and the 
stationary nucleation rate are the same for asymptotic approximations of both, the BDE and 
the ZFE. Thus our simplified theory (II) yields expressions for the nucleation rate that are 
similar to those found by Shneidman \cite{shn87} and by Shi et al \cite{shi90}, although 
their time lags differ from ours, as one would expect from Wu's arguments \cite{wu96}. 
For large critical sizes, our approximation (I) is better. 

The rest of the paper is as follows. In Section \ref{sec:model},
we review the Becker-D\"oring model for nucleation and growth of
spherical aggregates with the Turnbull-Fisher (TF) discrete diffusivity \cite{tur49}. 
The binding energy of the aggregate with $k$ monomers 
($k$ cluster) relative to isolated monomers in solution is $(k-1)$ times the
monomer-monomer bond energy plus a term proportional to the surface
area of the aggregate. Beyond a critical density no equilibrium size distribution
exists and the aggregates grow indefinitely. The main results of our asymptotic
analysis are derived in Section \ref{sec:asymptotics} and compared with the numerical 
solution of the BDE with the TF discrete diffusion coefficient describing 
devitrification of lithium disilicate glass. Our results compare favorably with previous
theories based on the ZFE, the continuum approximation of the BDE. However, when
compared with experimental data for glass disilicate, the theoretical time lag is about 30 
times smaller. To improve the agreement with experiments, we propose in Section 
\ref{sec:heat} a different discrete diffusion coefficient selected so as to yield the known 
expression for the adiabatic growth of a nucleus of critical size by heat transfer. The 
asymptotic theory for the resulting BDE is similar to that explained in Section 
\ref{sec:asymptotics}, and the resulting time lag is much closer to experimental data. 
Section \ref{sec:discusion} compares our asymptotic results for the transient nucleation 
rate and for the number of supercritical clusters to previously known analytical formulas 
(unfortunately all of them dealing with the continuum ZFE, not with the discrete BDE as 
ours do) \cite{tri87,shn87,shi90,shn91,dem93}. Technical matters are relegated to 
Appendices.

\section{Kinetic equations and stationary solutions}
\label{sec:model} 
The model presented here is nucleation in a lattice in which there are many more binding 
sites, $M$, than particles, $N$, \cite{neu02}. We shall consider the thermodynamic limit, 
$N\to\infty$ with fixed particle density per site, $\rho\equiv N/M$. Let $p_{k}$ be the
number of clusters with $k$ particles or, in short, $k$ clusters, and let $\rho_k\equiv 
p_k/M$ be the density of $k$ clusters. Note that the number densities per site, $\rho$ and 
$\rho_{k}$, are both dimensionless. Number densities per unit volume are obtained dividing 
$\rho$ and $\rho_{k}$ by the molecular volume, $v=V/M$. Particle conservation implies 
that the total particle density $\rho$ is constant:
\begin{eqnarray}
\sum_{k=1}^\infty k \rho_k = \rho . \label{e1}
\end{eqnarray}
In the Becker-D\"oring kinetic theory of nucleation, a $k$ cluster can grow or decay 
by capturing or shedding one monomer at a time. Then \cite{neu02}
\begin{eqnarray}
\dot{\rho}_{k} = j_{k-1}- j_k\equiv - D_-\, j_k,
\quad k\geq 2,    \label{e2}\\ 
j_{k} = d_{k}\,\left\{ e^{{D_+\varepsilon_{k}\over k_B T}}\, \rho_1 
\rho_k -  \rho_{k+1} \right\}.  \label{e3}
\end{eqnarray}
The monomer density $\rho_{1}$ can be obtained from the conservation identity 
(\ref{e1}) that relates it to the other cluster densities. In these equations, $\dot{\rho}_{k}
= d\rho_{k}/dt$ and $D_{\pm} u_k\equiv \pm [u_{k\pm 1}- u_{k}]$ are finite differences. 
$t$, $d_{k}$ and $j_{k}$ are nondimensional. $t$ and $d_{k}$ are related to the 
dimensional time $t^*$ and decay coefficient $d^*_{k}$ as follows
\begin{eqnarray}
t= \Omega t^*,\quad d_{k}= {d^*_k  \over\Omega}.  \label{e12}
\end{eqnarray}
Here the factor $\Omega$ has units of frequency, it depends on the particular model we 
choose for $d_{k}$, and will be determined later. $j_k$ is the net rate of creation of a $k+1$ 
cluster from a $k$ cluster (the {\em flux} in size space), given by the mass action law. 
In (\ref{e3}) we have made the detailed balance assumption to relate the kinetic coefficient 
for monomer aggregation to that of decay of a $(k+1)$ cluster, $d_{k}$. This implies that
the equilibrium size distribution solving $j_k =0$ has the form 
\begin{eqnarray} 
\tilde{\rho}_k = \rho_{1}^k\, \exp \left({\varepsilon_{k}\over k_{B}T}\right). 
\label{e4}
\end{eqnarray} 

In (\ref{e3}) and (\ref{e4}), $\varepsilon_{k}$ is the binding energy of a $k$ cluster, 
required to separate it into its monomer components. Then the total energy measured with 
respect to a configuration in which all clusters are monomers is $- \sum_{k=2}^{N} p_{k}
\varepsilon_{k}$. For spherical aggregates,
\begin{eqnarray}
\varepsilon_{k} = \left( (k-1) \alpha  - {3\over 2}\sigma
(k^{{2\over 3}} -1)\right)\, k_B T.  \label{e5}
\end{eqnarray}
This formula holds for $k\gg 1$, but we shall use it for all $k\geq 1$. $\alpha k_{B}T$
is the monomer-monomer bonding energy \cite{isr91} which, in the case of precipitation 
of crystals from a solution or segregation by coarsening of binary alloys, may depend on the 
particle density $\rho$ (volume fraction) through some empirical formulas \cite{pen83}. 
In Eq.\ (\ref{e5}), $\sigma=2\gamma_{s} (4\pi v^2/3)^{{1\over 3}}/(k_B T)$, 
where $\gamma_{s}$ and $v=V/M$ are the interfacial free energy per unit area (surface 
tension) and the molecular volume, respectively. Note that $\alpha$ and $\sigma$ are
both dimensionless. The correction $3\sigma k_{B}T/2$ in (\ref{e5}) ensures that 
$\varepsilon_{1}=0$, and it improves the agreement between the nucleation rate obtained 
from the BDE and experiments \cite{wu96}. More precise atomic models were proposed 
by Penrose et al \cite{pen83}.

Eqs.\ (\ref{e1}), (\ref{e2}), (\ref{e3}), (\ref{e5}) and a given discrete diffusivity
$d_{k}$ form a closed 
system of equations that we can solve for an appropriate initial condition. If initially only 
monomers are present, we have $\rho_{1}(0) = \rho$, and $\rho_{k}(0)=0$ for $k\geq 2$. 
Before we obtain formulas for the kinetic coefficient $d_{k}$, we shall recall the more 
salient features of the equilibrium size distribution.

\subsection{Equilibrium size distribution}
\begin{figure}
\begin{center}
\includegraphics[width=8cm]{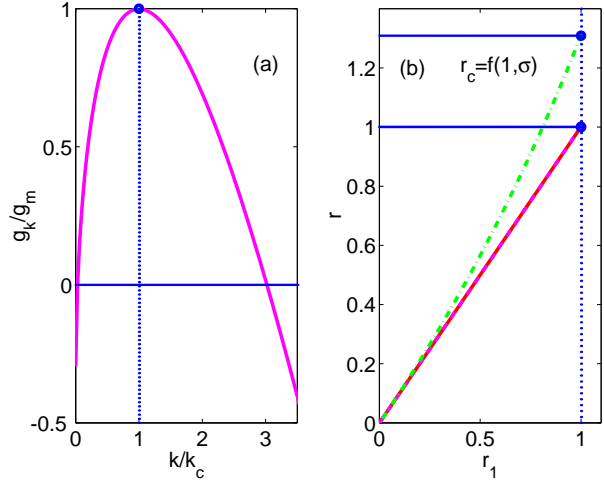}
\vspace{0.5 cm}
\caption{(a) Scaled activation energy $g_{k}/g_{m}$ as a function of the scaled 
size $k/k_c$. (b) Scaled dimensionless density $r=\rho e^\alpha$ as a function of the scaled 
dimensionless monomer density $r_1=\rho_1 e^\alpha$ for the equilibrium distribution 
(solid line). Data correspond to liquid iron at maximum undercooling (dot-dashed line), 
whereas for disilicate glass, $\rho\approx \rho_{1}$ (solid line). }
\label{fig1}
\end{center}
\end{figure}

The equilibrium distribution (\ref{e4}) satisfies $j_k =0$ and it can be written as
\begin{eqnarray}
&& \tilde{\rho}_k = \rho_{1}\, e^{-g_{k}}, \label{e6}\\
&& g_{k} \equiv - (k-1)\, \ln \rho_{1} - {\varepsilon_{k}\over k_{B}T}
=  {3\over 2}\sigma (k^{{2\over 3}}-1) - (k-1)\,\ln\left( e^\alpha 
\rho_{1}\right),  \label{e7}
\end{eqnarray}
where $g_{k}$ is the {\em activation energy}, equivalently given by
\begin{eqnarray}
g_{k} &=& \sigma_{k} - (k-1)\, \varphi, \quad
\sigma_{k} =  {3\over 2}\sigma\, (k^{{2\over 3}}-1),\quad (k\geq 1),\label{e8}\\
\varphi &=& \ln\left( e^\alpha \rho_{1}\right) .  \label{e9}
\end{eqnarray}
Here $\sigma_{1}= 0=g_{1}$. Assuming $k\gg 1$, $g_{k}$ achieves its global 
maximum $g_{m}= \sigma k_{c}^{2/3}/2 + \sigma k_{c}^{-1/3}-3\sigma/2$ at the 
critical size 
\begin{eqnarray} 
k = k_{c} \equiv \left({\sigma\over\varphi} \right)^3.  \label{e15}
\end{eqnarray}
Eq.\ (\ref{e8}) can be rewritten as
\begin{eqnarray} 
g_{k} \sim \sigma k^{2/3}_{c}\,\left\{ {3\over 2}\left({k\over k_{c}} 
\right)^{2/3} - {k\over k_{c}} \right\} + \sigma k_{c}^{-1/3}
-{3\sigma\over 2}.   \label{e16}
\end{eqnarray}
$g_{k}/g_{m}$ as a function of $k/k_{c}$ is depicted in Fig.~\ref{fig1}(a).  

Rewriting the flux (\ref{e3}) in the BDEs in terms of the activation energy, we obtain
\begin{eqnarray} 
j_{k} &=& d_{k}\, \left\{ \left(e^{- D_+ g_{k}} - 1\right)\, 
\rho_k - D_+ \rho_{k} \right\}.   \label{e13}
\end{eqnarray}
Eq.\ (\ref{e2}) is a spatially discrete Smoluchowski equation 
with diffusion coefficient $d_{k}$ and drift velocity
\begin{eqnarray} 
v_{k} = d_{k}\,\left(e^{- D_+ g_{k}} - 1\right).   \label{e14}
\end{eqnarray}
Notice that $v_{k}<0$ for an activation energy that increases with $k$ and $v_{k}>0$
for decreasing $g_{k}$. Hence, $g_{k}$ indicates how the discrete advection $v_{k}$
transports the clusters in size space: subcritical clusters shrink as time elapses while 
supercritical clusters grow with time. 

For the equilibrium densities (\ref{e6}), the conservation identity (\ref{e1}) becomes
\begin{eqnarray}
e^\alpha \rho = \sum_{k=1}^{\infty} k \left( e^{\alpha}\rho_{1}\right)^k\, 
e^{-\sigma_{k}} = \sum_{k=1}^{\infty} k\, e^{k\varphi-\sigma_{k}}. \label{e10}
\end{eqnarray}
This series converges for $e^{\alpha}\rho_{1}=e^\varphi \leq 1$ ($\varphi\leq 0$), 
and diverges for $e^{\alpha}\rho_{1} > 1$ ($\varphi>0$). At the critical micelle 
concentration (CMC), $\rho_{1} = e^{-\alpha}$ ($\varphi=0$), we obtain the critical 
density above which equilibrium is no longer possible,
\begin{eqnarray}
e^\alpha \rho_{c} = 1+ \sum_{k=2}^{\infty} k \, e^{-\sigma_{k}}. \label{e11}
\end{eqnarray}
For $\rho>\rho_{c}$, the BD kinetic equations predict phase segregation, i.e., indefinite
growth of ever larger clusters.

\subsection{The controlling parameters}
The simplest nucleation problem consists of solving the BD equations (\ref{e1}), 
(\ref{e2}) and (\ref{e13}), with dimensionless activation energy $g_{k}= \sigma_{k} 
- (k-1) \varphi$, discrete diffusivity $d_{k}$ (to be chosen later) and initial conditions
\begin{eqnarray}
\rho_1(0)=\rho,\,\, \rho_2(0) =\rho_3(0)=\ldots =0 . \label{e26}
\end{eqnarray}
The only parameters left in this initial value problem are $\rho$ and $\sigma$. $\rho$ 
controls the long-time behavior of the BDE: If $\rho\leq \rho_{c}$ given by (\ref{e11}), 
$\rho_{k}(t)$ approach their equilibrium values  (\ref{e6}), with 
monomer density $\rho_{1}$ that solves Equation (\ref{e1}): 
\begin{eqnarray}
\rho e^\alpha = f(\rho_{1}e^\alpha;\sigma) \equiv \sum_{k=1}^\infty\, k  (\rho_1
e^\alpha)^k e^{-\sigma_{k}}. \label{e27}
\end{eqnarray}
The graph of this function is either the solid line or the dashed line in Fig.~\ref{fig1}(b). 
If $\rho> \rho_{c}$, cluster sizes grow indefinitely whereas their density becomes 
small. Thus there remains a residual monomer concentration whose density $\rho_{1}
e^\alpha\to 1$ as $t\to\infty$. Summarizing, the union of solid or dashed lines in 
Fig.~\ref{fig1}(b) and the vertical line $\rho_{1}e^\alpha=1$ for $\rho> \rho_{c}$ 
represents the long-time limit of the monomer concentration as a function  of $\rho$. 

\begin{figure}
\begin{center}
\includegraphics[width=8cm]{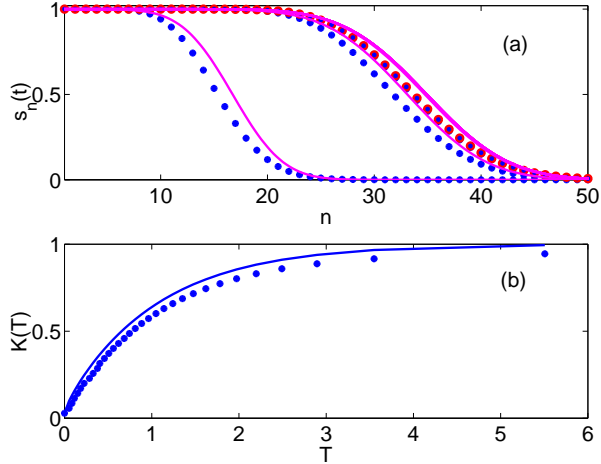}
\vspace{0.5 cm}
\caption{(a) Comparison of $s_{n}(t)$ evaluated (at different times) from the numerical 
solution of the discrete equations (\ref{e31}) to the asymptotic result (\ref{a23}) (solid line). 
(b) $K(T)$ calculated from Eq.\ (\ref{a9}) with $K(0)= \epsilon^3$ (solid line) is
compared to the numerically obtained position of the wave front. Data correspond to 
disilicate glass at 820 K. All variables are written in dimensionless units.}
\label{fig2}
\end{center}
\end{figure}

Let us identify the controlling parameters $\rho$ and $\sigma$ in a physical system
undergoing homogeneous nucleation. A good experimental example is the transformation 
of certain silicate glasses to crystals (devitrification) \cite{kel91}. In particular, abundant 
data exist for lithium disilicate and we have compiled in Table I appropriate values of 
parameters characterizing nucleation \cite{kel83}. In disilicate, the free energy per 
molecule of the crystal phase in the activation energy (\ref{e8}) is proportional to the 
undercooling
\begin{eqnarray}
\varphi = {\Delta S_{f} (T_{m}-T)\over N_{A}k_{B}T},  \label{g1}
\end{eqnarray}
where $T_{m}$ is the melting temperature, $\Delta S_{f}$ is the molar entropy of fusion 
and $N_{A}$ is Avogadro's number. The dimensionless density $\rho=e^{\varphi(0)}$ 
can be extracted from Eq.\ (\ref{g1}) as explained in Section \ref{sec:asymptotics}. In 
energy units, the activation free energy is $k_{B}T g_{k} = \gamma_{s} 4\pi a^{2} - 
k_{B}T \varphi k$, where $a$ is the radius of a spherical $k$ cluster. From the expression 
for the volume of this cluster, $kv = 4\pi a^3/3$ ($v$ is the molecular volume), we obtain 
$a = [3v/(4\pi)]^{1/3} k^{1/3}$, and therefore
\begin{eqnarray}
k_{B}T \,\left(g_{k} -\varphi + {3\sigma\over 2}\right) = 
\gamma_{s} (4\pi)^{1/3} (3v)^{2/3} k^{2/3} - \Delta S_{f}(T_{m}-T) k/N_{A}.
\label{g2}
\end{eqnarray}
Comparing (\ref{g2}) with (\ref{e8}) yields $\sigma = (32\pi v^2/3)^{1/3}\gamma_{s}
/(k_{B}T)$, and the critical size 
\begin{eqnarray}
k_{c}^{1/3}= \left({32 \pi v^2\over 3}\right)^{1/3} {\gamma_{s} N_{A}\over 
\Delta S_{f}\, (T_{m}-T)}. \label{g3}
\end{eqnarray} 
The other parameters in Table I will be used later to model the discrete 
diffusivity in the BDE. We observe that the critical size increases with temperature: $k_{c}= 
18$ at 703 K and $k_{c}= 34$ at 820 K. For other materials, such as undercooled liquid 
metals, critical sizes can be rather large: liquid iron at maximum undercooling has $k_{c}= 
494$, whereas $k_{c}= 2253$ for liquid rutenium at maximum undercooling \cite{kel91}.
\begin{table}[h]
\begin{center}
\begin{tabular}{*{3}{|c}|}    
\hline
Parameter & Symbol & Value\\
\hline
Melting temperature & $T_{m}$ & 1300 K\\
\hline
Entropy of fusion & $\Delta S_{f}$ & 40 J mol$^{-1}$ K$^{-1}$\\
\hline
Surface tension & $\gamma_{s}$  & 0.15 J/m$^2$\\
\hline
Preexponential diffusivity & $D_{0}$ & 2 $\times 10^9$ m$^2$ s$^{-1}$\\
\hline 
Activation energy for diffusion & $Q$ & 440 kJ/mol\\
\hline 
Molecular volume & $v$ & $10^{-28}$ m$^3$ \\
\hline 
TF time scale (703K) & $\Omega_{TF}^{-1}$  & 0.613 hours\\
 \hline
Heat capacity per unit volume & $\rho_{m}c$ & $10^6$ J m$^{-3}$ K$^{-1}$\\       \hline
Thermal conductivity (703K) & $\rho_{m}c\kappa$ & $3.96\times 10^{-18}$ J m$^{-1}$
s$^{-1}$ K$^{-1}$\\     \hline
Thermally-driven-growth time scale (703K) & $\Omega_{TDG}^{-1}$  & 6.196 hours\\
 \hline
Critical size (703K) & $k_{c}$ & 18 \\
\hline 
Undercooling (703K) & $\tilde{\varphi}$ & 4.087\\
\hline
Dimensionless surface tension (703K) & $\sigma$ & 10.74\\
\hline
Dimensionless free energy barrier (703K) & $g_{m}={\sigma\over 2} 
k_{c}^{2/3} -{3\sigma\over 2} + \tilde{\varphi}$ & 25.177\\    \hline
 TF time scale (820K) & $\Omega_{TF}^{-1}$  & 0.0478 s\\
 \hline
Thermal conductivity (820K) & $\rho_{m}c\kappa$ & $1.84\times 10^{-13}$ J m$^{-1}$
s$^{-1}$ K$^{-1}$\\   \hline
Thermally-driven-growth time scale (820K) & $\Omega_{TDG}^{-1}$  & 0.48 s\\
 \hline
Critical size (820K) & $k_{c}$ & 34 \\
\hline 
Undercooling (820K) & $\tilde{\varphi}$ & 2.817\\
\hline
Dimensionless surface tension (820K) & $\sigma$ & 9.207\\
\hline
Dimensionless free energy barrier (820K) & $g_{m}={\sigma\over 2} 
k_{c}^{2/3} -{3\sigma\over 2} + \tilde{\varphi}$  & 38.181\\    \hline
\end{tabular}
\label{tab1}
\end{center}
\caption{Data for lithium disilicate glass}
\end{table}

\subsection{Equivalent Becker-D\"oring system}
As they stand, the BDE are rather stiff and hard to solve numerically. For example, at 
equilibrium, Table I indicates that $\rho_{k_{c}}/\rho_{1} = e^{-g_{m}-\tilde{
\varphi}}\approx e^{-25.2-4.1} \approx 2 \times 10^{-13}$ for disilicate glass at 703 K, 
and $\rho_{k_{c}}/\rho_{1} \approx 1.6 \times 10^{-18}$ for disilicate glass at 820 K. 
This motivates the following change of variable
\begin{eqnarray}
\rho_{k} = \rho_{1} e^{-g_{k}} s_{k} = e^{-\alpha} e^{k\varphi-\sigma_{k}} s_{k}, 
\label{e28}
\end{eqnarray}
according to (\ref{e9}). Note that $s_{k}=1$ in equilibrium. Since $g_{1}=0$, this 
equation implies 
\begin{eqnarray}
s_{1}\equiv 1,    \label{e29}
\end{eqnarray}
for all $t$. For the initial condition (\ref{e26}), $e^{\varphi(0)-\alpha}=\rho_{1}(0) = 
\rho$, and the conservation identity (\ref{e1}) becomes
\begin{eqnarray}
e^{\varphi(0)} = e^{\varphi} + \sum_{k=2}^\infty\, k\, 
 e^{k\varphi -\sigma_{k}} s_{k}, \label{e30}
\end{eqnarray}
in which we have used (\ref{e28}). In terms of the $s_{k}$, the flux can be written as 
\begin{eqnarray}
j_{k}= d_{k} \exp[(k+1)\varphi - \sigma_{k+1}]\, (s_{k}- s_{k+1}), \label{flux}
\end{eqnarray}
and the BDE (\ref{e2}) and (\ref{e13}) become
\begin{eqnarray}
\dot{s_{k}} + u_{k} (s_{k+1} - s_{k}) = -k \dot{\varphi} s_{k}
+ d_{k-1}\, (s_{k-1} - 2s_{k} + s_{k+1}), \label{e31}
\end{eqnarray}
for $k\geq 2$. Here, 
\begin{eqnarray}
u_{k} = d_{k-1} - d_{k} e^{\varphi - D_{+}\sigma_{k}}. \label{e32}
\end{eqnarray}
The term $u_{k}\, D_{+}s_{k}$ in Eq.\ (\ref{e31}) represents {\em discrete 
advection}, with a drift velocity $u_{k} = - v_{k} +(d_{k-1} - d_{k})\sim - v_{k}$, which
is essentially minus the drift velocity in the original BDE for $k\gg 1$. Thus, the advection 
 in Eq.\ (\ref{e31}) {\em climbs up} the activation energy barrier, from small values of 
 $g_{k}$ to large ones.
 
 In summary, the transformed nucleation initial-boundary value problem consists of the 
 balance equations (\ref{e31}), the particle conservation equation (\ref{e30}), the 
 boundary condition (\ref{e29}), $s_{1}=1$, and initial conditions $s_{k}(0)=0$ for all 
 $k\geq 2$. Its solution gives $\varphi(t)$ and $s_{k}(t)$ for all $k\geq 2$ and all $t>0$.
 
  \subsection{Stationary solution}
The stationary solution of the BDE has a flux independent of cluster size, so that  $j_{k}= 
d_{k} \exp[(k+1)\varphi - \sigma_{k+1}]\, (s_{k}- s_{k+1})= j$, from which $(s_{k+1}
- s_{k}) = - j \exp[\sigma_{k+1} - (k+1)\varphi]/d_{k}$, and therefore
 \begin{eqnarray}
s_{k}= 1-j \sum_{l=1}^{k-1} {\exp[\sigma_{l+1}-(l+1)\varphi]\over d_{l}},
\label{e33}
\end{eqnarray}
for $k\geq 2$. Since $s_{\infty}=0$, $j$ can be obtained from this expression in terms
of an infinite series
\begin{eqnarray}
j= {1\over\sum_{l=1}^{\infty} \exp[\sigma_{l+1}-(l+1)\varphi -\ln d_{l}]}.
\label{e34}
\end{eqnarray}
Substituting back this expression into (\ref{e33}), we obtain
 \begin{eqnarray}
s_{k}= 1- {\sum_{l=1}^{k-1} \exp[\sigma_{l+1}-(l+1)\varphi -\ln d_{l}]\over 
\sum_{l=1}^{\infty} \exp[\sigma_{l+1}-(l+1)\varphi -\ln d_{l}]}.
\label{e35}
\end{eqnarray}
Then, $\rho_{k} = \rho_{1}\, e^{-g_{k}} s_{k}$. 

\subsection{Turnbull-Fisher discrete diffusivity}
To solve the BDE, we need to establish reasonable models of the kinetic coefficient $d_{k}$ 
(discrete diffusivity) for the decay of the $(k+1)$ cluster. A classical formula due to Turnbull
and Fisher \cite{tur49} applies to spherical clusters whose growth is limited by the reaction 
rate at their boundary: $d^*_{k}$ is the product of the number of active sites on the 
aggregate times the molecular jump rate \cite{tur49,kel83}
\begin{eqnarray}
d^*_{k}=  4\, k^{2/3}\, e^{D_{+}g_{k}/2} {6 D\over \lambda^2}
= \Omega\, k^{2/3}\, e^{D_{+}g_{k}/2},\quad \Omega^{-1}= {v^{2/3}\over 24 D} 
\equiv {v^{2/3}e^{Q/(RT)}\over 24 D_{0}}.  \label{e17}
\end{eqnarray}
Here $D=D_{0} e^{-Q/(RT)}$ is the diffusion coefficient in the liquid, $Q$ is the
activation energy for diffusion (see Table I), $R=k_{B}N_{A}$ is the gas constant and
$\lambda = v^{1/3}$ ($v$ is the molecular volume). If we nondimensionalize time
as in Eq.\ (\ref{e12}) with this definition of $\Omega$, we obtain
\begin{eqnarray}
d_{k}=  k^{2/3}\, e^{D_{+}g_{k}/2} .  \label{e18}
\end{eqnarray}

\section{Asymptotic theory of transient homogeneous nucleation with the Turnbull-Fisher
diffusivity}
\label{sec:asymptotics}
In this section, we shall interpret the numerical solutions shown in Figures \ref{fig2} to
\ref{fig4} by using singular perturbation methods. Our theory will be described using the 
TF discrete diffusivity (\ref{e17}) and compared to numerical solution of the 
BDE for the crystallization of disilicate glass at different undercoolings. 

\subsection{Initial transient}
Initially, $\rho_1(0) = \rho$ and there are no multiparticle aggregates. There is an 
initial transient stage during which dimers, trimers, etc.\ form at the expense of the 
monomers. This initial stage is characterized by the decay of the chemical driving force 
$\varphi=\alpha+\ln \rho_{1}$ to a quasi-stationary value $\tilde{\varphi}$, given 
by Eq.\ (\ref{g1}) in the case of disilicate glass, and the emergence of a continuum size 
distribution. Knowing this, {\em we choose the initial chemical driving force $\varphi(0)$ so
that the quasistationary value $\tilde{\varphi}$ given by Eq.\ (\ref{g1}) is attained
at the end of the initial stage.}

In materials such as disilicate glass at the temperatures we consider, the
critical size is relatively small. Then $\varphi(0)\approx\tilde{\varphi}$, and the 
initial stage is very short. As the critical size increases (as in the case of undercooled liquid
metals), $\varphi(0)$ may differ appreciably from $\tilde{\varphi}$, and the initial 
stage lasts longer. However, even in such cases, the duration of the initial stage, $t_{
\infty}$, is negligible if we are interested in the overall duration of the transient stage to 
quasi-stationary nucleation. We shall show later that the duration of the initial stage compared 
to the duration of the overall transient is of order $k_{c}^{-2/3}$, a very small quantity for 
materials with large critical sizes.

\subsection{Wave front advancing towards the cluster of critical size}
After the first era, clusters of increasing size are formed. For sufficiently small clusters, 
the continuum size distribution approaches the equilibrium distribution with $\varphi=
\tilde{\varphi}$. This situation can be observed as an advancing wave front in the variable 
$s_{k}(t)$, satisfying $s_{k}\sim 1$ (equilibrium) behind the front and $s_{k}\sim 0$
ahead of the front. This second era is described by Equations (\ref{e30}) to (\ref{e32})
with $\varphi=\tilde{\varphi}$ and $\dot{\varphi}=0$. The critical sizes,
\begin{eqnarray} 
k_{c} = \left({\sigma\over\tilde{\varphi}}\right)^3 , \label{a1}
\end{eqnarray}  
 for disilicate glass are relatively small, between 10 and 50, but they are large for
undercooled liquid metals, generally between 100 and 1000. Hence we shall use as a small 
gauge parameter 
\begin{eqnarray} 
\epsilon = {\tilde{\varphi}\over\sigma}. \label{a2}
\end{eqnarray}  
Our asymptotic analysis will be carried out in the limit $\epsilon\to 0$, and therefore 
$k_{c}= \epsilon^{-3}\to\infty$. Then $d_{k}$, $u_{k}$ and $\sigma_{k}$ in Eqs.\ 
(\ref{e18}), (\ref{e31}) and (\ref{e32}) are smooth functions of $k>0$:
\begin{eqnarray} 
d(k) = k^{2/3}\, e^{[D_{+}\sigma(k) - \tilde{\varphi}]/2},\quad 
\sigma(k) = {3\over 2} \sigma\, (k^{2/3} -1),   \label{a3}\\
u(k) = d(k-1) - d(k)\, \exp[\tilde{\varphi} - \sigma(k+1) + \sigma(k)]. \label{a4}
\end{eqnarray}  

\subsubsection{Position of the wave front}
In the numerical solutions shown in Fig.~\ref{fig2}(a), the graphs of  $s_{k}$ vs.\ $k$ at 
fixed time have clear inflection points at some $k$, where $s_{k}\approx 1/2$. The 
inflection point is taken as the {\em position} of the wave front. In the continuum model, 
the front position $k= k_{f}(t)$ is a smooth function which obeys
\begin{eqnarray}
\dot{k}_{f} = u(k_{f}).   \label{a5}
\end{eqnarray}  
Scaling $k_{f}$ as 
\begin{eqnarray}
k_{f} = {K\over\epsilon^3}  \label{a6}
\end{eqnarray}  
(same scaling as $k_{c}= \epsilon^{-3}$), the right hand side of (\ref{a5}) becomes
\begin{eqnarray} 
u(k_f) &=& {1\over \epsilon^2}\, U(K) + O(\epsilon), \label{a7}\\
U(K) &=& 2 K^{2/3}\, \sinh\left( {\tilde{\varphi}\over 2} (K^{-1/3}-1)
\right) .  \label{a8}
\end{eqnarray}  
Eq.\ (\ref{a5}) can be rewritten as 
\begin{eqnarray}
{dK\over dT} = U(K) \equiv 2 K^{2/3}\, \sinh\left( {\tilde{\varphi}\over 2}
(K^{-1/3}-1) \right) ,   \label{a9}
\end{eqnarray}  
provided we define the slowly varying time scale $T=\epsilon t$, and take the limit as
$\epsilon\to 0$. Fig.~\ref{fig2}(b) compares the position of the wave front calculated
by solving (\ref{a9}) with $K(0)=\epsilon^3$ to the value obtained from the numerical
solution of (\ref{e31}). Note that the solution of (\ref{a9}) presents a time shift with 
respect to the numerical solution of the discrete model. This time shift reflects the breakdown 
of the continuum limit as $K\to 0$, due to discreteness, and also the transient in $\varphi(t)$ 
before it settles to $\tilde{\varphi}$. If the solution of Eq.\ (\ref{a9}) is forced to agree 
with the numerical $K(T)$ when the latter is, say, 0.1, the comparison fares much better.

\subsubsection{Shape of the wave front}
The leading edge of the wave front is a layer centered at $K(T)$ in which $s_k$ decreases 
from 1 to 0 as $k$ increases through it. The continuum representation of $s_k$ in this layer is 
\begin{eqnarray}
s_{k} = S(X,T;\epsilon), \label{a11}
\end{eqnarray}
where is $S$ a smooth function of its arguments and $X$ is the scaled displacement
from the wave front location at $k=K/\epsilon^3$, i.e.,
\begin{eqnarray}
X = \epsilon^p \left( k - {K\over \epsilon^3} \right).   \label{a12}
\end{eqnarray}
The scaling exponent $p$, presumably with $0<p<3$, is to be determined. The description 
(\ref{a11}) - (\ref{a12}) should hold as $\epsilon \to 0$ with $X$ fixed, so that the layer
thickness scales as $\epsilon^{-p}$. Substituting (\ref{a11}) into (\ref{e31}) yields, 
\begin{eqnarray} 
\epsilon {\partial S\over\partial T} - \epsilon^{p-2} {dK\over dT}{\partial
S\over\partial X} + u\left( {K\over\epsilon^3} + {X\over\epsilon^p}\right) \,
[S(X+\epsilon^p,T;\epsilon) - S(X,T;\epsilon)]\nonumber\\
 = d\left( {K\over\epsilon^3} + {X\over\epsilon^p}\right) \,
[S(X-\epsilon^p,T;\epsilon) - 2 S(X,T;\epsilon) + S(X+\epsilon^p,T;\epsilon) ] .
\label{a13}
\end{eqnarray}
Carrying out the straightforward expansion in powers of $\epsilon$,  (\ref{a13}) 
adopts the following asymptotic form
\begin{eqnarray} 
&&\epsilon {\partial S\over\partial T} + \epsilon^{p-2} \left[ U(K) - 
{dK\over dT}\right]\, {\partial S\over\partial X} + \epsilon U'(K) X
{\partial S\over\partial X} \nonumber\\
&& = \epsilon^{2p-2} \left[K^{2/3} e^{\tilde{\varphi}\, (K^{-1/3}-1)/2} - 
{1\over 2}U(K)\right] \, {\partial^2 S\over\partial X^2} + o(\epsilon^{2p-2}),
\label{a14}
\end{eqnarray}
as $\epsilon\to 0$ with $X$, $K$ fixed. Here $U(K)$ is given by (\ref{a8}). To obtain
(\ref{a14}), we have used (\ref{a6}) and (\ref{a7}): 
$$u\left(K+\epsilon^{3-p}X\over\epsilon^{3}\right) = 
\epsilon^{-2} U(K+\epsilon^{3-p}X) + O(\epsilon) = \epsilon^{-2} U(K) +
\epsilon^{1-p} U'(K)\, X + o(\epsilon^{1-p}).$$
The dominant balance of diffusion and 
convection in (\ref{a14}) yields $2p-2=1$, or $p=3/2$. Hence (\ref{a12}) yields
\begin{eqnarray}
X = \epsilon^{3/2} \left( k - {K\over \epsilon^3} \right),   \label{a15}
\end{eqnarray}
and the limit of (\ref{a14}) as $\epsilon\to 0$ is 
\begin{eqnarray}
&& {\partial S\over\partial T} + U'(K) X {\partial S\over\partial X} 
= D(K)\, {\partial^2 S\over\partial X^2} , \label{a16}\\
&& D(K)\equiv \lim_{\epsilon\to 0}\left[d(\epsilon^{-3}K)- {1\over 2}\,
u(\epsilon^{-3}K)\right] \epsilon^2  = K^{2/3} \cosh\left( {\tilde{\varphi}
\over 2}(K^{-1/3}-1) \right). \label{a17}
\end{eqnarray}
Had we carried out the same analysis for the ZFE, we would have found $D(K)\sim d(
\epsilon^{-3}K)\,\epsilon$. This would have resulted in a wider wave front and a longer
time to ignition than those described below.

\subsubsection{Flux and wave front width}
Besides determining the shape of the wave front near its location, Eq.\ (\ref{a16}) 
yields the behavior of the flux (creation rate of clusters larger than $k$) $j_{k}$ near $k=
k_{f}$. If we substitute (\ref{e18}), (\ref{a11}) and (\ref{a15}) into (\ref{flux}):
\begin{eqnarray} 
j_{k} &=& - d_{k} e^{(k+1)\tilde{\varphi}-\sigma_{k+1}} D_{+}s_{k} \nonumber\\
&=& - k^{2/3} \exp\left[
\left(k+{1\over 2}\right)\tilde{\varphi} - {3\sigma\over 4}[(k+1)^{2/3}+
k^{2/3}]+{3\sigma\over 2} \right]\, D_{+}s_{k},   \nonumber 
\end{eqnarray} 
we obtain
\begin{eqnarray} 
j_k \sim \epsilon^{-1/2}\, K^{2/3} \, e^{3\tilde{\varphi}/(2\epsilon)} 
\exp\left[- {G(K)\over\epsilon^3} - {G'(K)X\over\epsilon^{3/2}}
- {G'(K)\over 2} - {G''(K)\over 2}\, X^2\right]\, {\partial S\over\partial X}.\,  
\label{a18}
\end{eqnarray}  
Here 
\begin{eqnarray}
G(K) \equiv \tilde{\varphi}\, \left({3\over 2} K^{2/3} - K\right)  \label{a19}
\end{eqnarray}
is a scaled version of the activation energy (\ref{e8}). 

Since $j_k$ is proportional to $\partial S/\partial X$, it is convenient to differentiate
(\ref{a16}) with respect to $X$ in order to obtain an equation for $J\equiv - 
\partial S/\partial X$,
\begin{eqnarray} 
{\partial J\over\partial T} + U'(K)\, {\partial (X\, J)\over\partial X} = 
D(K)\, {\partial^2 J\over\partial X^2}. \label{a20}
\end{eqnarray}  
Notice that $J$ is locally conserved, and the following integral conservation identity holds:
\begin{eqnarray}
1 = - [S]_{-\infty}^\infty = - \int_{-\infty}^\infty {\partial S\over\partial X}
dX = \int_{-\infty}^\infty J\, dX. \label{a21}
\end{eqnarray}
Eq.\ (\ref{a20}) has Gaussian solutions satisfying (\ref{a21}), 
\begin{eqnarray}
J(X,T) = {1\over 2\sqrt{\pi A(T)}}\,\exp\left[ -{X^2\over 4\, A(T)}\right],  
\label{a22}
\end{eqnarray}
which yields 
\begin{eqnarray}
S(X,T) = {1\over 2}\,\mbox{erfc}\left[ {X\over 2\sqrt{A(T)}}\right] \label{a23}
\end{eqnarray}
for the wave front profile. Inserting Eq.\ (\ref{a22}) in Eq.\ (\ref{a20}), we find the 
following equation for $A(T)>0$:
\begin{eqnarray}
{dA\over dT} &-& 2\, U'(K)\, A = D(K). \label{a24} 
\end{eqnarray}  
Since $K(T)$ is an increasing function, we can express $A$ as a function of $K$. Inserting
(\ref{a9}) in (\ref{a24}), we get
\begin{eqnarray}
{dA\over dK} &-& {2\, U'(K)\over U(K)}\, A = {D(K)\over U(K)}.
\end{eqnarray}  
Direct integration of this equation yields
\begin{eqnarray} 
A= q U^2 + U^2\int D {dK\over U^3} = {3 K^{4/3}\left(1+q\sinh^2[{
\tilde{\varphi}(K^{-1/3}-1)\over 2}]\right)\over 2\tilde{\varphi}} , 
\label{A-eq}
\end{eqnarray}
in which $q$ is an arbitrary constant.

After insertion of (\ref{a22}), the flux (\ref{a18}) becomes
\begin{eqnarray} 
j_k \sim {K^{2/3}\, e^{3\tilde{\varphi}/(2\epsilon)}\over 2\sqrt{\epsilon\pi 
A}}\, \exp\left\{ - {G(K)\over\epsilon^3} - {G'(K)X\over\epsilon^{3/2}}
-{G'(K)\over 2} -\left[{G''(K)\over 2}+ {1\over 4A}\right]\, X^2\right\}.  
\label{a25} 
\end{eqnarray}  
Here $K=K(T)$ and $A=A(T)$ are found by solving the differential equations (\ref{a9}) 
and (\ref{a24}) with initial conditions $K(0)=\epsilon^3$ and $A(0)= 3\epsilon^4/(2
\tilde{\varphi})$, respectively. We have to set $q=0$ in Eq.\ (\ref{A-eq}) for 
$A$ would become exponentially large as $K=\epsilon^3\to 0$ otherwise. As $T\to
\infty$, $K\to 1$ and $A\to - D(1)/[2U'(1)]$. The definitions (\ref{a8}) and (\ref{a17}) 
of $U(K)$ and $D(K)$ imply $U'(1)= -\tilde{\varphi}/3$, $D(1)= 1$. Hence, $A\to 3/(2 
\tilde{\varphi})$ as $T\to\infty$, or as $K\to 1$ in Eq.\ (\ref{A-eq}). The definition 
(\ref{a19}) of $G(K)$ implies $G(1) = \tilde{\varphi}/2$, $G'(1)=0$, and $G''(1) = -
\tilde{\varphi}/3$. Hence, the limit as $T\to\infty$ of the creation rate (\ref{a25}) is
\begin{eqnarray} 
j_k \sim j_{\infty}\equiv \sqrt{{\tilde{\varphi}\over 6\pi\epsilon}}\,
\exp\left(- {\tilde{\varphi}\over 2\epsilon^3} + {3\tilde{\varphi}\over 2
\epsilon}\right).    \label{a27} 
\end{eqnarray}
Notice that the terms proportional to $X$ and $X^2$ have disappeared from this expression 
and therefore  $j_{\infty}$ is asymptotically uniform for $X=O(1)$. Eq.\ (\ref{a27}) is
the classical quasi-steady nucleation rate of supercritical clusters due to Zeldovich 
\cite{zel43}, and it can be directly obtained from the stationary flux (\ref{e34}) in the 
limit as $\epsilon\to 0$. 

\subsection{The nucleation rate of supercritical clusters}
Let us now study the transient creation rate, in which $j\equiv j_{k_{c}}$ increases from 0 
to the steady Zeldovich value (\ref{a27}). As we have just seen, our theory predicts that 
{\em the wave front profile is given by (\ref{a23}), where $K(T)$ and $A(T)$ are solutions 
of Eqs.\ (\ref{a9}) and (\ref{a24}), respectively. The flux of 
clusters with sizes larger than $k$ is then given by Eq.\ (\ref{a25})}. Setting $k=k_{c}=
\epsilon^{-3}$ (critical size) and $X=(1-K(T))/\epsilon^{3/2}$ in this equation, we 
obtain the nucleation rate predicted by our theory, $j(t)$. Its integral over time yields the 
number of supercritical clusters, $N_{c}(t)$. We shall consider now a different and more 
explicit approximation of these results. 

\begin{figure}
\begin{center}
\includegraphics[width=8cm]{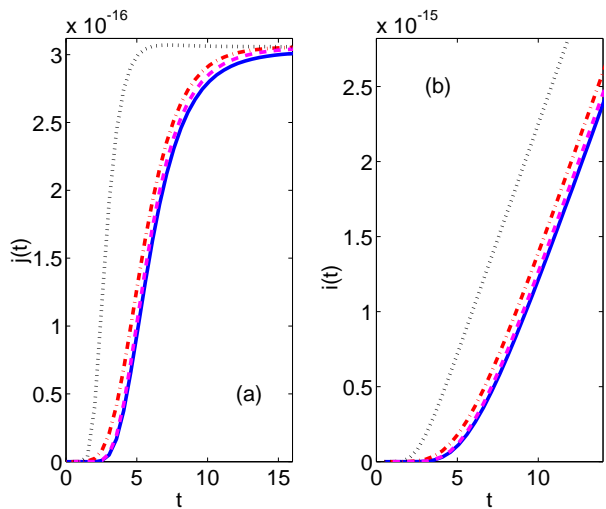} 
\vspace{0.5 cm}
\caption{(a) Evolution of the dimensionless flux at critical size $j(t)$, and (b) number 
of clusters surpassing critical size $N_{c}(t)$ as a function of dimensionless time
for disilicate glass at 820K, $k_{c}=34$. Solid lines correspond to numerical results, dashed 
lines to the approximation given by Eq.\ (\ref{a25}), dot-dashed lines to the linearization 
approximation (\ref{a32}) and dotted lines to the approximation (\ref{l8}) corresponding 
to linearizing the equations for $K(T)$ and $A(T)$ as in Appendix \ref{sec:ap4}. }
\label{fig3}
\end{center}
\end{figure} 

\subsubsection{Linearization of the wave front speed about the critical size}
Let us fix $k=k_{c}=\epsilon^{-3}$ (critical size) in the definition (\ref{a15}) of $X$:
\begin{eqnarray} 
X = {1-K\over \epsilon^{3/2}} \equiv \kappa.    \label{a28} 
\end{eqnarray}
We now set $X=\kappa$ in (\ref{a25}) and perform the limit as $\epsilon\to 0$ with
$\kappa$ fixed. The result is
\begin{eqnarray} 
j \sim  j_{\infty} e^{- \tilde{\varphi}\kappa^2/6- \epsilon^{3/2}
\tilde{\varphi}\kappa/6} \sim  j_{\infty} e^{- \tilde{\varphi}\kappa^2/6},  
 \label{a29} 
\end{eqnarray}  
provided we use the limiting stationary value $(4A)^{-1} = - G''(1)/2$.

The transient turns on when $\kappa\equiv (1-K)/\epsilon^{3/2} = O(1)$. Since 
$U(1-\epsilon^{3/2}\kappa) \sim \epsilon^{3/2}\tilde{\varphi}\kappa/3$, 
the wave front equation (\ref{a8}) yields
\begin{eqnarray} 
{d\kappa\over dT} = - {\tilde{\varphi}\over 3}\,\kappa,     \label{a30} 
\end{eqnarray}
as $\epsilon\to 0$. The solution of this equation is 
\begin{eqnarray} 
\kappa &=& \kappa_{M}\, e^{- \tilde{\varphi} e^{\tilde{\varphi}}(T-T_{M})/3}
= \kappa_{M} e^{-(t-t_{M})/(2\tau)},     \label{a42}\\  
\tau^{-1} &=& {2\over 3} \tilde{\varphi} \epsilon. \label{a40}
\end{eqnarray}
It is convenient to choose $\kappa_{M}$ as the value of $\kappa$ at which the flux $j$ 
reaches its inflection point. Then we may consider that the wave front has ignited the 
nucleation of supercritical clusters. Straightforward use of Eqs.\ (\ref{a29}) and 
(\ref{a30}) shows that 
 \begin{eqnarray}
\kappa_{M} =\sqrt{{6\over\tilde{\varphi}}}.      \label{a41}
\end{eqnarray}
Moreover, $T_{M} = \epsilon t_{M}$ is the {\em time to ignition}, at which the wave 
front $K(T)$ reaches the value $K=1-\epsilon^{3/2}\kappa_{M}$. From (\ref{a9}), we
obtain
\begin{eqnarray} 
&& t_{M} = t_{\infty} + {3 \over 2 \tilde{\varphi}\epsilon}
\left\{ \ln\left({\tilde{\varphi}(1-\epsilon^3)^2\over 6 \epsilon^3}\right) 
\right.\nonumber\\
&&Ê\quad + \left.
\int_{\epsilon^3}^{1-\epsilon^{3/2}\kappa_{M}} \left[{\tilde{\varphi}\over 
3 K^{2/3}\sinh\left[{\tilde{\varphi}\over 2}(K^{-1/3} -1)\right] } 
+ {2\over K-1}\right] dK\right\} ,  \label{a34} 
\end{eqnarray}  
where $t_{\infty}$ is the duration of the initial stage. We could have expanded the 
integral in this expression, but Eq.\ (\ref{a34}) is better suited for numerical calculation. 
The nucleation rate is found by inserting Eq.\ (\ref{a42}) in (\ref{a29}):
\begin{eqnarray} 
j \sim j_{\infty} \exp\left[-e^{- (t-t_{M})/\tau}\right],  \label{a32}  
\end{eqnarray}  
in which a term of order $\epsilon^{3/2}$ has been ignored in the exponential.

Integrating $j(t)$ over time, we find the number of supercritical clusters as a function of 
time. In the limit as $t\to\infty$, this number is $N_{c}(t)\sim j_{\infty}\,( t-\theta)$, 
where the time lag $\theta$ is approximately given by $\theta= t_{M} + \tau
\gamma+ \tau E_{1}(e^{t_{M}/\tau})$, or
\begin{eqnarray} 
&&\theta= t_{\infty} + {3 \over 2 \tilde{\varphi}\epsilon}
\left\{\ln\left({\tilde{\varphi}(1-\epsilon^3)^{2}\over 6 \epsilon^3 }\right) 
+ \gamma +\tau E_{1}(e^{t_{M}/\tau})\right.\nonumber\\
&& \quad + \left.\int_{\epsilon^3}^{1-\epsilon^{3/2}\kappa_{M}} 
\left[{\tilde{\varphi}\over 3 K^{2/3}\sinh\left[{\tilde{\varphi}\over 
2}(K^{-1/3} -1)\right] } + {2\over K-1}\right] dK\right\}, \quad 
\label{a36}
\end{eqnarray}  
where $\gamma = 0.577215\ldots$ is Euler's constant and $E_{1}(x)$ is an exponential 
integral, see the derivation in Appendix \ref{sec:ap1}. The time lag $\theta$ can be 
directly compared to experimental values \cite{kel91}. 

\begin{figure}
\begin{center}
\includegraphics[width=9cm]{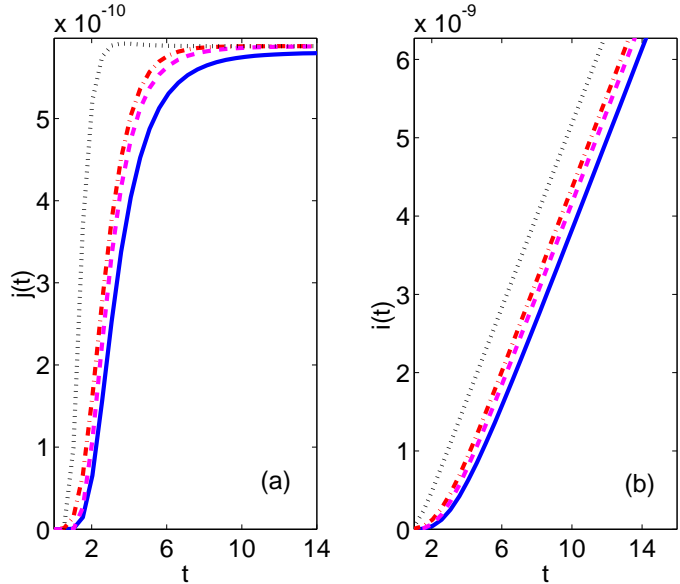}
\vspace{0.5 cm}
\caption{Same as in Fig.~\ref{fig3} for disilicate glass at 703K, $k_{c}=18$. }
\label{fig4}
\end{center}
\end{figure} 

\subsubsection{Comparison between different approximations}
Fig.~\ref{fig3}(a) compares $j(t)$ calculated from the numerical solution of the BDE for 
devitrification of disilicate glass at 820 K, from (\ref{a32}) and (\ref{a34}) with 
$t_{\infty}=0$, and from Eq.\ (\ref{a25}) with $X=(1-K(T))/\epsilon^{3/2}$. We 
find that the more precise expression, Eq.\ (\ref{a25}), captures better the width and 
location of the transition region between $j=0$ and $j=j_{\infty}$, as compared with the 
simple approximation given by Eqs.\ (\ref{a32}) and (\ref{a34}). Both approximations
present a small overshoot and yield a smaller time lag $\theta$ than that obtained from
the numerical solution of the BDE. The overshoot decreases as the critical size 
increases. Another approximation consists of linearizing the equations for $K(T)$ and
$A(T)$ about the critical size $K=1$ as suggested in Ref. \cite{dem93} and further
explained in Appendix \ref{sec:ap4}. This latter approximation is the worst one. This is not 
surprising as such approximation provides the same result for both the discrete BDE and the
continuum ZFE.

For disilicate glass at a lower temperature of 703 K, the critical size is smaller and
our approximations deviate more from the numerical solution of the BDE, as shown in 
Figure \ref{fig4}(a). Integrating $j(T)$ over time, we find the number of supercritical 
clusters as a function of time, $N_{c}(t)$, which is depicted in Figs. \ref{fig3}(b) and 
\ref{fig4}(b). At 703 K, the numerical solution of the BDE with the TF diffusivity yields 
a time lag $\theta= 2.6$. This value is close to those provided by the linearization 
approximation, $\theta=2.2$, and by Eq.\ (\ref{a25}), $\theta=2.3$. Thus these
analytical approximations to the numerical solution are reasonably good even for a
relatively small critical size. However, $\theta=2.6$ gives 1.6 hours according to 
Table I, whereas the experimentally measured time lag is about 50 hours, cf.\ Fig.\ 5 
of Ref.~\cite{kel91}. This discrepancy is due to having used the TF discrete 
diffusivity, which yields an excessively small time unit, as shown in Table I.

\section{Temperature driven growth of the nucleus}
\label{sec:heat}
To improve agreement with experiments, we need a discrete diffusivity different
from the TF one. We shall no longer assume that cluster size changes due to the activated
transfer of a monomer through the cluster surface as in the TF theory. Instead, we shall 
assume that the discrete diffusivity in the BDE agrees with an adiabatic temperature 
driven growth of the nucleus. This yields a different formula for $d_{k}$ which,
presumably, is not physically justified for very small cluster sizes. Nevertheless, the 
numerical solution of the BDE corresponding to thermally driven growth provides a time lag 
which is much closer to the experimentally measured value for disilicate glass than the TF
diffusivity.

\subsection{Discrete diffusivity}
Let us assume that there is a nonuniform temperature field about a spherical crystal of radius 
$a$ ($k$ cluster) that is growing at the expense of the surrounding glass. Eq.\ (\ref{g3}) 
shows that a nucleus of critical size grows if $(T_{m}-T)$ decreases. The same equation 
yields $(T_{m}-T)$ at the surface of a critical nucleus with $k$ monomers: 
\begin{eqnarray}
T_{m}-T_{*} = \left({32\pi v^2\over 3}\right)^{1/3} {\gamma_{s} N_{A}\over
\Delta S_{f}}\, k^{-1/3} .  \label{g4}
\end{eqnarray}
At the surface of the crystal, $T=T_{*}(k)$, whereas far from it there is a smaller 
temperature, $T=T_{\infty}$. Heat transfer from crystal to glass, $-\rho_{m} c\kappa 
4\pi a^2\, \partial T(a,t)/\partial r$ ($\kappa$, $c$ and $\rho_{m}$ are thermal 
diffusivity, specific heat and mass density, respectively), should equal the increase of energy 
due to crystal growth, $T_{\infty} \Delta S_{f}\, N_{A}^{-1}dk/dt^*$. We find
\begin{eqnarray}
\left. {dk\over dt^*}= -{\rho_{m} c\kappa N_{A}\over T_{\infty} 
\Delta S_{f}}  {\partial T\over\partial r}\right|_{r=a}\, 4\pi a^2.   \label{g5}
\end{eqnarray}
The temperature $T(r)$ is the solution of Laplace's equation with boundary conditions $T= 
T_{*}$ at $r=a$, and $T= T_{\infty}$ infinitely far from the nucleus. The corresponding 
solution is  $T=T_{\infty} + (T_{*}-T_{\infty})a/r$, which, together with (\ref{g5}), 
yield
\begin{eqnarray}
{dk\over dt^*} = 4\pi {\rho_{m} c\kappa N_{A}\over T_{\infty} 
\Delta S_{f}}\, (T_{*}-T_{\infty})\, a.     \label{g6}
\end{eqnarray}
Using Eq.\ (\ref{g4}) and the expressions for the radius $a$ and (\ref{g3}) for the critical 
size $k_{c}$, we obtain 
\begin{eqnarray}
{dk\over dt^*}= {2 (6\pi^2 v)^{1/3} \rho_{m} c\kappa N_{A}(T_{m}
-T_{\infty})\over T_{\infty} \Delta S_{f}}\, (k^{1/3} - k_{c}^{1/3}).   \label{g7}
\end{eqnarray}
As $k\to\infty$, the flux (\ref{e13}) in the BDE becomes $j^*_{k}\sim 
d^*_{k}\, (e^{-\partial g_{k}/\partial k} - 1)\,\rho_{k}$ (written in 
dimensional units), and therefore 
\begin{eqnarray}
{dk\over dt^*}\sim  d^*_{k}\, (e^{\varphi-\sigma k^{-1/3}}-1) = 
d^*_{k}\, e^{-\sigma k^{-1/3}} \left(e^{\varphi}-e^{\sigma k^{-1/3}}
\right) ,  \nonumber
\end{eqnarray}
Using Eq.\ (\ref{e15}), this equation can be written as 
\begin{eqnarray}
{dk\over dt^*}\sim d^*_{k}\,\left(e^{\varphi\, [1- (k_{c}/k)^{1/3}]}
-1 \right).     \label{e24}
\end{eqnarray}
Comparing Eqs.\ (\ref{g7}) and (\ref{e24}), we obtain
\begin{eqnarray}
d^*_{k}= \Omega k^{1/3}{\varphi\, \left[1- \left({k_{c}\over k}
\right)^{1/3}\right]\over e^{\varphi\, [1- (k_{c}/k)^{1/3}]} -1}, \quad \quad 
\Omega =  {2 (6\pi^2 v)^{1/3} \rho_{m} c\kappa
k_{B}N_{A}^2\over (\Delta S_{f})^2} .   \label{g8}
\end{eqnarray}
As before, we shall absorb the constant $\Omega$ in the definition of time according to 
(\ref{e12}), which yields the following value of the dimensionless discrete diffusivity:
\begin{eqnarray}
d_{k}= k^{1/3}{\varphi\, \left[\left({k_{c}\over k}\right)^{1/3} - 1
\right]\over 1 - e^{\varphi\, [1- (k_{c}/k)^{1/3}]} }.   \label{g8a}
\end{eqnarray}
Here we shall assume that the thermal diffusivity follows the same Arrhenius law as the 
diffusion coefficient in the liquid $\kappa = D_{0} e^{Q/(N_{A}k_{B}T)}$; see Kelton et 
al \cite{kel83}. With this choice of discrete diffusivity, a numerical solution of the BDE
yields a time lag of $\theta=46.5$ hours at 703 K compared to $\theta=1.6$ hours 
previously obtained using the TF diffusivity. The experimentally measured time 
lag is 50 hours, as shown in Fig.\ 5, page 94 of Kelton's review \cite{kel91}. Thus we 
feel justified in using our formula (\ref{g8}) to solve the BDE for disilicate glass.

\subsection{Asymptotic theory}
We have to repeat the arguments given in Section \ref{sec:asymptotics} using the discrete 
diffusivity (\ref{g8}) instead of the TF expression. One important difference is
that time needs to be rescaled as $T=\epsilon^2 t$ instead of $T=\epsilon t$. Here we
shall also use the symbol $T$ for the slow time scale, but remembering that $T_{TDG}=
\epsilon^2 t$ (for thermally driven growth) instead of $T_{TF}=\epsilon t$ (TF 
diffusivity). When necessary, we shall add the labels TF or TDG to the corresponding 
variables. After the initial discrete stage, our asymptotic theory yields the following results 
for temperature driven cluster growth, applicable to devitrification of disilicate glass:
\begin{itemize}
\item The wave front profile $S(X,T)$, with $X=[k-\epsilon^{-3}K(T)]\epsilon^{3/2}$ 
and $T=\epsilon^2 t$, is given by $S(X,T)=(1/2)$erfc$(X/[2\sqrt{A(T)}])$. The 
front location and its width solve
\begin{eqnarray} 
{dK\over dT} &=& U(K) \equiv \tilde{\varphi}\, (1-K^{1/3}),   \label{g9} \\
 {dA\over dT} &-& 2 U'(K)\, A = D(K) \equiv {\tilde{\varphi}\over 2}\, 
 (1-K^{1/3}) \coth\left[{\tilde{\varphi}\over 2} (K^{-1/3}-1) \right],\label{g10}
\end{eqnarray}  
with initial conditions $K(0)=\epsilon^3$, $A(0)= \epsilon^3/2$. The latter condition
corresponds to $q=0$ in Eq.\ (\ref{A-eq}). Then $A\sim K/2$ as $K\to 0+$, which
yields the initial condition for $A$ if $K(0)=\epsilon^3$. In Eqs.\ (\ref{g9}) and 
(\ref{g10}), $U(K)$ and $D(K)$ are defined by
$$ U(K) = \lim_{\epsilon\to 0} \left[\epsilon\,\, u\left({K\over\epsilon^3}
\right)\right], \quad 
D(K) = \lim_{\epsilon\to 0}\left[d\left({K\over\epsilon^3}\right) 
- {1\over 2} u\left({K\over\epsilon^3}\right)\right]\, \epsilon.
$$
Instead of Eq.\ (\ref{a18}), we get the following approximation for the flux near the wave
front:
\begin{eqnarray} 
j_k \sim {\epsilon^{1/2}\tilde{\varphi}\, (1-K^{1/3}) \, e^{3\tilde{\varphi}/(2
\epsilon)}\over [1- e^{-G'(K)}]\sqrt{4\pi A}}\, \exp\left\{- {G(K)\over
\epsilon^3} - {G'(K)X\over\epsilon^{3/2}} \right. \nonumber\\
\left. - G'(K) - \left[{G''(K)\over 2}+ {1\over 4A}\right] X^2\right\}\,,     
\label{g12}
\end{eqnarray}  
in which $G(K)= \tilde{\varphi}\, (3K^{2/3}/2 - K)$ and Eq.\ (\ref{a22}) has been 
used. Inserting $X=[1-K(T)]/\epsilon^{3/2}$ in this equation, we obtain the nucleation rate: 
\begin{eqnarray} 
&& {j(T)\over j_{\infty}} \sim \sqrt{{3\over 2A\tilde{\varphi} }} {U(K)\over
1-e^{-G'(K)}}\, 
\exp\left\{ {\tilde{\varphi}\over 2\epsilon^3} - {G(K)+G'(K)(1+\epsilon^3-K)
\over\epsilon^{3}} \right. \nonumber\\
&& \quad \quad \quad \left. - \left[{G''(K)\over 2}+ {1\over 4A}\right]\, 
{(1-K)^2\over\epsilon^3}\right\},   \quad   \label{g11} \\
&& j_{\infty}=\sqrt{{\epsilon\tilde{\varphi}\over 6\pi}}\,\exp\left({3
\tilde{\varphi}\over 2\epsilon} - {\tilde{\varphi}\over 2\epsilon^3}\right). 
\label{g13}
\end{eqnarray}  
\item The simplest approximation for the nucleation rate and the time lag yields 
\begin{eqnarray} 
j &=& j_{\infty}\,\exp\left\{ - e^{-(t-t_{M})/\tau} \right\},
 \label{g15}\\
t_{M} &=& {1\over \tilde{\varphi}\epsilon^2}\int_{\epsilon^3}^{1-
\epsilon^{3/2}\kappa_{M}} {dK\over 1-K^{1/3}} \nonumber\\
 &=& {3\over 2\tilde{\varphi}\epsilon^2}\left\{ \ln\left({3
 \tilde{\varphi}\over 2\epsilon^3}\right) - 3 + 2 \left(
 {6\epsilon^3\over\tilde{\varphi}}\right)^{1/2}\right\} + O(\epsilon),
 \label{g16a}\\
\tau_{TDG}^{-1} &=& {2\tilde{\varphi}\epsilon^2\over 3} = \epsilon
\tau_{TF}^{-1}.     \label{g17}\\
\theta &=& t_{M} + [\gamma + E_{1}(e^{t_{M}/\tau})]\, \tau\nonumber\\
&=& {3\over 2\tilde{\varphi}\epsilon^2}\left\{ \ln\left({3
 \tilde{\varphi}\over 2\epsilon^3}\right) - 3 + \gamma + 2 \left(
 {6\epsilon^3\over\tilde{\varphi}}\right)^{1/2}\right\} + O(\epsilon),
 \label{g16}
\end{eqnarray}  
To obtain these expressions, we have followed the same procedure as in the case of the 
TF diffusivity. In particular, Eqs.\ (\ref{a29}) - (\ref{a42}) hold with $T=\epsilon^2 t$,
$t_{M}$ given by Eq.\ (\ref{g16a}) and $\tau$ given by Eq.\ (\ref{g17}).
\end{itemize}
\begin{figure}
\begin{center}
\includegraphics[width=9cm]{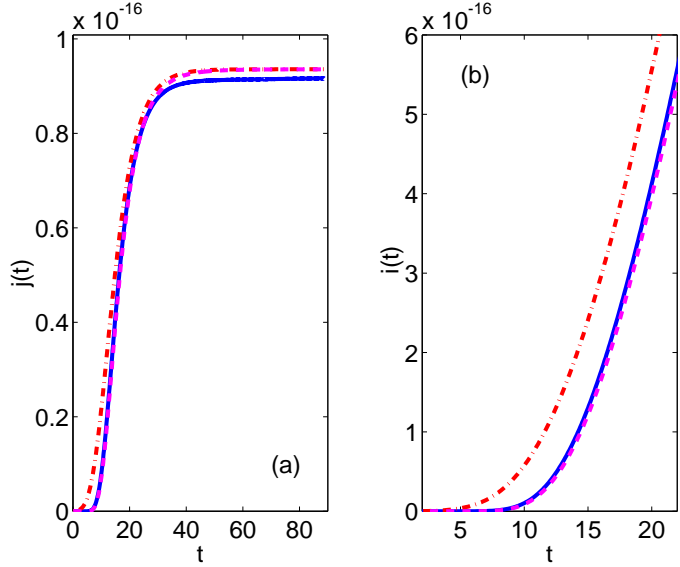}
\vspace{0.5 cm}
\caption{(a) Evolution of the dimensionless flux at critical size $j(t)$, and (b) number 
of clusters surpassing critical size $N_{c}(t)$ as a function of time (in dimensionless units)
for disilicate glass at 820K, $k_{c}=34$. Solid lines correspond to numerical results, dashed 
lines to the approximation given by Eqs.\ (\ref{g9}) to (\ref{g11}), and dot-dashed lines 
to Eqs.\ (\ref{g15}) - (\ref{g17}). }
\label{fig5}
\end{center}
\end{figure} 

\begin{figure}
\begin{center}
\includegraphics[width=9cm]{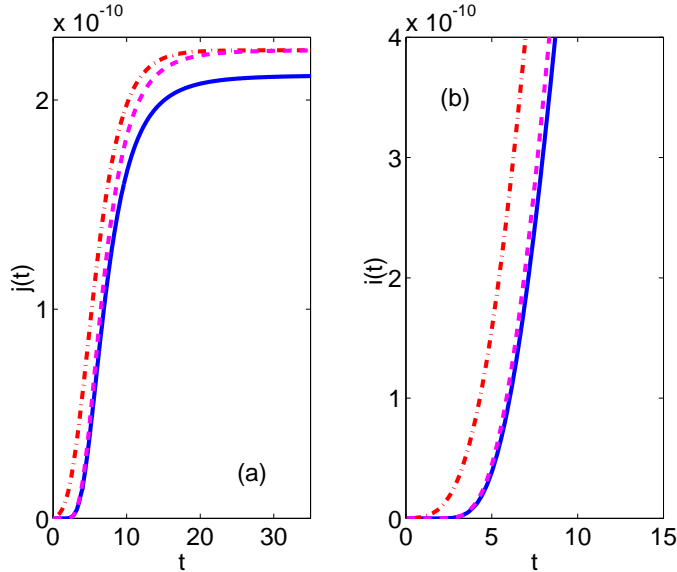}
\vspace{0.5 cm}
\caption{Same as Figure \ref{fig5} for disilicate glass at 703K, which has a critical size
 $k_{c}=18$. }
\label{fig6}
\end{center}
\end{figure} 

Figure \ref{fig5} compares the numerical solution of the BDE (solid line) for devitrification 
of disilicate glass at 820 K (critical size $k_{c}=34$) with the more accurate asymptotic 
formulas: Eqs.\ (\ref{g15}) - (\ref{g17}) (dot-dashed line) and its linearization about
the critical size, Eqs.\ (\ref{g9}) to (\ref{g11}) (dashed line). Similarly, Fig.~\ref{fig6} 
corresponds to 703K. We observe that our two approximations, Eqs.\ (\ref{g9}) - 
(\ref{g11}), and Eq.\ (\ref{g15}), describe quite accurately the numerical solution. 
Notice that our asymptotic formulas for thermally driven growth yield worse approximations 
to the numerical solution of the BDE than in the case of the TF diffusivity. The stationary 
nucleation rate is approximated less well by $j_{\infty}$ in the case of thermally driven 
growth because of the avoidable singularity of $d_{k}$ at the integer $k_{c}$, which is 
slightly different from $\epsilon^{-3}$.

\section{Discusion}
\label{sec:discusion}
In this paper, we have studied the case of phase segregation resulting when $\rho>\rho_{c}$.
Previously, other authors had carried out asymptotic studies of the BDE in the simpler 
case of subcritical density, $\rho<\rho_{c}$, in which initial conditions of only monomers, 
or more general ones, evolve towards the equilibrium distribution. In many cases of 
polynomial growth for $d_{k}$, equilibrium is reached via a wave front profile for $s_{k}$, 
which is similar to Eq.\ (\ref{a23}) with $A\propto K^\delta$, and $K\propto T^\mu$, 
for appropriate positive $\delta$ and $\mu$; see Ref. ~\cite{kin02} and references cited 
therein. This advancing and widening wave front leaves in its wake the equilibrium size 
distribution. 

In the more complex case of phase segregation and indefinite aggregate growth considered in 
this paper, a quasicontinuum wave front of $s_{k}$ emerges after a short transient which is
governed by the discrete BDE. After this, the leading edge of the wave front advances 
towards the critical size, and it slows down and stops there, leaving behind it a 
quasi-equilibrium state. The arrival of the wave front to the critical size marks the {\em 
ignition} of nucleation of supercritical clusters, which ends when the stationary Zeldovich 
rate is reached. Previous asymptotic theories have been derived for the continuum ZFE, not 
the discrete BDE, and thus their results systematically misrepresent two things: (i) the time 
lags for transient nucleation, as explained by Wu \cite{wu96}, and (ii) the width of the 
wave front and the time to ignition in the nucleation rate. The latter discrepancies occur 
because the diffusion coefficient appearing in the continuum equation for the wave front 
satisfies $D_{BDE}(K) = D_{ZFE}(K) - U(K)/2$, and therefore the time to ignition in the 
nucleation rate for the BDE is {\em smaller} than the corresponding one for the ZFE. 

Let us briefly mention several existing asymptotic theories for the ZFE. Shneidman 
\cite{shn87} and Shi et al \cite{shi90} Laplace transformed 
the continuum ZFE and matched a first stage of pure advection of clusters to a local 
expansion about the wave front when it is near its final position at the critical size. They 
obtained our simplest formula for the nucleation rate, Eq.\ (\ref{a32}) with the same 
relaxation time, $\tau_{TF}$ or $\tau_{TDG}$, except that their values for $t_{M}$ were 
different from (\ref{a34}). This can be expected from Wu's arguments about approximating
the discrete BDE by the continuum ZFE \cite{wu96}; see the systematic shift of 
approximations of the ZFE with respect to numerical solutions of the BDE in  Fig.\ 20 of 
Ref.~\cite{wu96}. Trinkaus and Yoo \cite{tri87} studied a ZFE with a drift term 
linearized about the critical size (parabolic barrier) as an approximation to the full ZFE. 
Their results are comparable to those found by means of the Laplace transform and matched 
asymptotic expansions; see Wu's review \cite{wu96}. All these authors 
obtained a transition region for the nucleation rate $j(t)$ that was 
wider than observed in the numerical solution of the BDE. Several authors also found
a nucleation rate for supercritical clusters that did not tend to $j_{\infty}$ as $t\to\infty$
if $k\neq k_{c}$ \cite{tri87,shi90,dem93}, which is often called the {\em asymptotics 
catastrophe} \cite{mak00}. Our theory is free from this deficiency: Eq.\ (\ref{b2}) in 
Appendix \ref{sec:ap1} provides the flux at $k>k_{c}$ using the TF diffusivity
\begin{eqnarray} 
j_{k} &=& j_{\infty}\, e^{-\tilde{\varphi}X_{0}\kappa/3}\, 
 e^{-\tilde{\varphi}\kappa^2/6}\nonumber\\
 &=&  j_{\infty}\, \exp\left[- X_{0}\sqrt{2\tilde{\varphi}/3}\,
 e^{-(t-t_{M})/(2\tau)}\right]\,  e^{- e^{-(t-t_{M})/\tau}},   \label{c1}
\end{eqnarray}
in which $X_{0}= \epsilon^{3/2}\, (k-k_{c})$. Notice that $j_{k} \sim j_{\infty}$ as 
$t\to\infty$, even after making our simplest approximation: linearization of the wave
front about the critical size. To get rid of the asymptotics catastrophe, Maksimov et al
\cite{mak00} assumed that $S(X,T)= (1/2)$ erfc$\{[A e^{-t/(2\tau)} + B(X)]/\sqrt{
1-\zeta e^{-t/\tau}}\}$, in which the new function $B(X)$ obeyed an ad hoc self-consistent 
equation that ensured $j_{k}\sim j_{\infty}$ as $t\to\infty$ even if $k\neq k_{c}$. Note
that if we use (\ref{a42}) for $X=\kappa$ and the linearization approximation for $A$ as
in Appendix \ref{sec:ap4}, we obtain the previous formula for $S$ with $\zeta=1$, 
$A=e^{t_{M}/(2\tau)}$ and $B=0$. Shneidman \cite{shn01} criticized Maksimov et al's 
result and extended his earlier asymptotic formula for the nucleation rate \cite{shn91} to 
non-critical sizes. The previous criticism of using approximations to the ZFE instead of 
approximations to the discrete BDE apply to these works. Our more precise 
approximation using Eq.\ (\ref{a25}) plus the exact equations for the wave front location 
and its instantaneous width improve upon these approximations and perform better for 
materials with large critical sizes.

The time lag obtained from the numerical solution of the BDE with the TF diffusivity
(or from our asymptotic approximations using it) is too small as compared with experimental 
results (about thirty times smaller for disilicate at 703 K). The TF discrete diffusivity
yields an excessively small time unit, as shown in Table I. We have greatly improved the 
agreement of theory and experiments by using a different formula for the discrete diffusivity,
which is found by imposing that the growth rate of a critical nucleus resulting from the BDE 
be the same as obtained by heat transfer. In this case, our asymptotic approximations have 
a slightly different scaling of time and different expressions for $U(K)$ and $D(K)$.

\acknowledgments 
Part of this work was carried out during a visit of J.\ Neu's to the Universidad Carlos III de 
Madrid, whose support we acknowledge. The present work was financed by the Spanish 
MCyT grants BFM2002-04127-C02-01 and BFM2002-04127-C02-02, and by the European 
Union under grant HPRN-CT-2002-00282. 

\appendix
\section{General solution of Eq.\ (\ref{a20}) }
\label{sec:ap0}
It is convenient to rewrite this equation in terms of the variables $K$ and $X$, as
\begin{eqnarray} 
{\partial J\over\partial K} + {U'(K)\over U(K)}\, {\partial (X\, J)\over\partial X} = 
{D(K)\over U(K)}\, {\partial^2 J\over\partial X^2}, \label{ap1}
\end{eqnarray}  
to be solved with the homogeneous boundary condition 
\begin{eqnarray} 
\left(\epsilon^{-3/2} + {U'(K)X_{in}\over U(K)}\right)\, J - {D(K)\over U(K)}\, 
{\partial J\over\partial X} = 0 \label{ap3}
\end{eqnarray}  
(at $X=X_{in}\equiv \epsilon^{3/2} - K \epsilon^{-3/2}$, corresponding to $k=1$ in the
definition of $X$), and with initial condition $J(X,K_{0})= - \partial S_{0}(X)/\partial 
X$. The boundary condition is obtained by differentiating 
\begin{eqnarray} 
S(X_{in},T)=1 
\end{eqnarray}
with respect to $T$ and then using the definition of $J$ and Eqs.\ (\ref{a9}) and 
(\ref{a16}). The solution of the initial-boundary value problem is 
\begin{eqnarray} 
 J(X,K) = -\int_{-\infty}^\infty G(X,K;X_{0},K_{0})\, {\partial
 S_{0}(X_{0})\over\partial X_{0}}\, dX_{0} , \label{ap2}
\end{eqnarray}  
where the Green's function $G(X,K;X_{0},K_{0})$ satisfies Eq.\ (\ref{ap1}) with initial 
condition $G(X,K_{0}+;X_{0},K_{0}) = \delta(X-X_{0})$ and the same homogeneous
boundary condition as $J$ at $X=X_{in}$. In a simple application of the method of images,
the Green's function for this BVP can be written in terms of the Green's function 
$G_{\infty}(X,K;X_{0},K_{0})$ for the infinite real line $X$, as 
\begin{eqnarray}
 G(X,K;X_{0},K_{0}) &=& G_{\infty}(X,K;X_{0},K_{0}) \nonumber\\
&+&  c(X_{0};K,K_{0})\, G_{\infty}(X,K;2X_{in}U_{0}/U - X_{0},K_{0}), 
\label{ap4}\\
 c(X_{0};K,K_{0}) &=&{ {\left({X_{in}\over U} - {X_{0}\over U_{0}}\right)D
 \over \left(\epsilon^{-3/2} + {U' X_{in}\over U}\right)2UB} +1 \over {\left(
 {X_{in}\over U} - {X_{0}\over U_{0}}\right)D\over 
 \left(\epsilon^{-3/2} + {U' X_{in}\over U}\right)2UB} -1} , \label{ap5}\\
  B(K,K_{0}) &=& \int_{K_{0}}^{K} {D(K)\, dK\over U(K)^3} .  \label{green2}
\end{eqnarray}
Now, $G_{\infty}(X,K;X_{0},K_{0})$ can be calculated by first writing an equation for 
the Fourier transform $\hat{G}_{\infty}(\xi,K;X_{0},K_{0}) = \int_{-\infty
}^\infty e^{i\xi X} G_{\infty}(X,K;X_{0},K_{0}) dX$. Such an equation is a 
first-order quasilinear hyperbolic equation that can be solved by the method of characteristics. The result is that 
$\hat{G}_{\infty}$ is Gaussian in $\xi$. Inverting the Fourier transform, we obtain 
\begin{eqnarray}
G_{\infty}(X,K;X_{0},K_{0}) &=& 
{e^{ -{ \left({X\over U}- {X_{0}\over U_{0}}\right)^2\over 4 B(K,K_{0})}
}\over U\sqrt{4\pi B(K,K_{0})}} .  \label{green1}
\end{eqnarray}
Given the initial condition $S_{0}(X_{0})= H(2\epsilon^{3/2}-K_{0}\epsilon^{-3/2} - 
X_{0})$ (pure monomers), Eq.\ (\ref{ap2}) yields 
\begin{eqnarray}
J(X,K) = G(X,K;2\epsilon^{3/2}- \epsilon^{-3/2}K_{0},K_{0}) \sim 
{e^{ -{ X^2\over 4 U^2\, B(K,K_{0})}}\over U\sqrt{4\pi B(K,K_{0})}} ,  
\label{green3}
\end{eqnarray}
which is Eq.\ (\ref{a22}), up to exponentially small terms. Here $K_{0}=\epsilon^3$, 
$U_{0} = U(K_{0}) = 2\epsilon^2 \sinh[\tilde{\varphi}/(2\epsilon) - 
\tilde{\varphi}/2]\sim \epsilon^2 e^{\tilde{\varphi}/(2\epsilon) - 
\tilde{\varphi}/2}$, $U = U(K)$, and $K= K(T)$.  

\section{Calculating the time lag}
\label{sec:ap1}
The time $T_{M}$ can be estimated from Eq.\ (\ref{a9}) with initial condition $K(0)=
\epsilon^3$ (pure monomers) as
\begin{eqnarray} 
&& T_{M} - \epsilon t_{\infty} = \int_{\epsilon^3}^{1-\epsilon^{3/2}\kappa_{M}}
{dK\over U(K)}  \nonumber\\
&& \quad = \int_{\epsilon^3}^{1-\epsilon^{3/2}\kappa_{M}}{dK\over U'(1) (K-1)} 
+ \int_{\epsilon^3}^{1-\epsilon^{3/2}\kappa_{M}}
\left[{1\over U(K)} -{1\over U'(1) (K-1)}\right] dK ,\quad  \label{app1} 
\end{eqnarray}  
where $t_{\infty}$ is the duration of the initial discrete stage in the original time scale. 
After straightforward calculations, we obtain (\ref{a34}).

The number of supercritical clusters is
\begin{eqnarray} 
N_{c} &\sim&Ê\int_{0}^{t} j(t) dt  = j_{\infty}\left\{t +Ê\int_{0}^{t}  \left[
\exp\left(-e^{- (t-t_{M})/\tau}\right) -1\right]\, dt \right\}
\sim j_{\infty}\, (t - \theta),  \label{app2} \\
\theta &\equiv& Ê\int_{0}^{\infty}  \left[ 1-
\exp\left(-e^{- (t-t_{M})/\tau}\right) \right]\, dt = \tauÊ
\int_{0}^{e^{t_{M}/\tau}}  {1- e^{-x}\over x}\, dx \nonumber\\
&=& t_{M} +\tau\gamma + \tau\, E_{1}\left(e^{t_{M}/\tau}
\right), \quad\quad   \label{app3}
\end{eqnarray}  
where $E_{1}(x)$ is an exponential integral and $\gamma = 0.577215\ldots$ is Euler's 
constant \cite{abr65}. Notice that $\tau\, E_{1}(x)\sim \tau e^{-x}/x\sim 9 
\epsilon [\tilde{\varphi}(1-\epsilon^3)]^{-2} e^{-\tilde{\varphi}/(6\epsilon^3)}
\ll\epsilon\ll 1$, as $x = e^{t_{M}/\tau}\sim \tilde{\varphi}/(6\epsilon^3)
\gg 1$ \cite{abr65}. Thus we can ignore the exponential integral in (\ref{app3}), which 
then yields Eq.\ (\ref{a36}). 

Sometimes it is interesting to calculate the creation rate of clusters of size $k>k_{c}$. If $k$
is close to critical size, we can write 
\begin{eqnarray} 
X = \epsilon^{3/2}\left( k - {K\over\epsilon^3}\right) = \epsilon^{3/2}\, (k - 
\epsilon^{-3}) + \kappa.    \label{b1} 
\end{eqnarray}
Thus $X= X_{0}+\kappa$, with $X_{0}= \epsilon^{3/2}\, (k - \epsilon^{-3})$.
Inserting $X= X_{0}+\kappa$ and $K=1-\epsilon^{3/2}\kappa$ in (\ref{a25}), we 
obtain the creation rate of clusters of size $k=\epsilon^{-3} + X_{0}\epsilon^{-3/2}$:
\begin{eqnarray} 
j_{k} &=& j_{\infty}\, e^{-\tilde{\varphi}X_{0}\kappa/3}\, 
 e^{-\tilde{\varphi}\kappa^2/6}\nonumber\\
 &=&  j_{\infty}\, \exp\left[- X_{0}\sqrt{2\tilde{\varphi}/3}\,
 e^{-(t-t_{M})/(2\tau)}\right]\,  e^{- e^{-(t-t_{M})/\tau}}.   \label{b2}
\end{eqnarray}
Notice that $j_{k} \sim j_{\infty}$ as $t\to\infty$. Thus our asymptotic result for
the flux over any cluster size is free from the {\em asymptotics catastrophe} \cite{mak00}: 
several authors found that their expressions for the flux tend to $J\neq j_{\infty}$ as $t\to
\infty$ if $k\neq k_{c}$. These catastrophes are due to inappropriate assumptions they
made in their derivations. 

\section{Linearization of the equations for $K(T)$ and $A(T)$ about the critical size}
\label{sec:ap4}
A possible approximation of the wave front equations consists of linearizing the equations 
for $K(T)$ and $A(T)$ about the critical size $K=1$:
\begin{eqnarray} 
{dK\over dT}\approx U'(1) (K-1),            \label{l1}\\
 {dA\over dT} - 2 U'(1)\, A\approx D(1),    \label{l2} 
\end{eqnarray}
with $K(0)= \epsilon^3$ and $A(0)\approx 0$ (initial condition of pure monomers). Both
for the TF discrete diffusivity and for temperature driven growth of the nucleus, we have
\begin{eqnarray} 
 U'(1) = - {\tilde{\varphi}\over 3}, \quad\quad D(1) = 1.    \label{l6} 
\end{eqnarray}
The solutions of Equations (\ref{l1}) and (\ref{l2}) can be written in terms of the 
time scale $t$ as
\begin{eqnarray} 
K(t) \approx 1 - (1-\epsilon^3) \,e^{-t/(2\tau)} ,    \label{l3}\\
A(t) \approx {3\over 2\tilde{\varphi}}\left(1 - e^{- t/\tau}\right),   
\label{l4} 
\end{eqnarray}
for the TF diffusivity, and the same formulas with $\tau_{TDG}$ instead of $\tau_{TF}$ 
for thermally driven growth. Near the critical size, these equations would give an explicit 
expression of the wave front profile (\ref{a23}), with $X= \epsilon^{3/2} k - 
\epsilon^{-3/2} K(T)$. The nucleation rate of supercritical clusters is then obtained from 
(\ref{a25}) with $X= (1 - K)/\epsilon^{3/2}$, together with (\ref{l3}) and (\ref{l4}). 
For the TF diffusivity, $j \equiv j_{k_{c}}$ is 
\begin{eqnarray} 
{j \over j_{\infty}} \sim {\left[1- (1-\epsilon^3) e^{-t/(2\tau)}
\right]^{2/3}\over\sqrt{1- e^{-t/\tau} } }\, \exp \left\{ -\tilde{\varphi}
\left[{\Psi(t)\over\epsilon^3} + {\left(1- (1-\epsilon^3) e^{-t/(2\tau)}
\right)^{-1/3}-1\over 2}\right] \right\} ,  \label{l7}\\
\Psi(t) = \left[1- (1-\epsilon^3) e^{-t/(2\tau)}\right]^{2/3} + 
\left[1- (1-\epsilon^3) e^{-t/(2\tau)}\right]^{-1/3} - 2 \nonumber\\
+ {(1-\epsilon^3)^2 e^{- t/\tau}\over 6} \left\{ {1\over 1- e^{-t/\tau}} 
- \left[1- (1-\epsilon^3) e^{-t/(2\tau)}\right]^{-4/3}\right\}.    \label{l5} 
\end{eqnarray}  

Notice that the same results would have been obtained from the ZFE because the difference
between our $D(K)$ and the corresponding one for the ZFE is $U(K)/2$, which vanishes at
$K=1$. If we replace 1 instead of $[1- (1-\epsilon^3) e^{-t/(2\tau)}]$ in the previous 
formulas, we find:
\begin{eqnarray} 
j \sim {j_{\infty}\over\sqrt{1- e^{-t/\tau}}}\, \exp \left\{ -
{\tilde{\varphi}\, (1-\epsilon^3)^2\over 6 \epsilon^3} \, 
{e^{-2t/\tau}\over 1- e^{-t/\tau}} \right\}.    \label{l8} 
\end{eqnarray}  
Demo and Koz\'\i sek's theory for the ZFE \cite{dem93} would yield Eq.\ (\ref{l8})
for the nucleation rate once a couple of errors are corrected. They found $\tau_{DK} 
= 7\tau_{TF}/5$ instead of the correct relaxation time $\tau_{TF}$, and an extra factor 
of 9 in the argument of the exponential in Eq.\ (\ref{l8}). Moreover, their exponential 
contains a factor $(1-\epsilon)$ instead of $(1-\epsilon^3)$. Demo and Koz\'\i sek's 
figure 3 shows that their formulas do not improve as the cluster size increases, as one would 
expect of correct asymptotic expressions. Instead, they seem to optimize the nucleation rate 
of crystals in disilicate at $k_{c}=27$ (T=800K), as compared with numerical solutions. 
The earlier theory by Trinkaus and Yoo \cite{tri87} calculated the Green function for a 
time-dependent ZFE with a quadratic barrier and also used a linear equation for the position 
of the wave front. Thus their results are related to those in this Appendix.


\begin{thebibliography}{99}
%
\bibitem[*]{neu:email} {E-address \tt neu@math.berkeley.edu}.
\bibitem[$\dag$]{bonilla:email} {E-address \tt bonilla@ing.uc3m.es}.
Author to whom all correspondence should be addressed.
\bibitem[$\ddag$]{carpio:email} {E-address \tt ana$_{-}$carpio@mat.ucm.es}.
\bibitem{ll10} E.M. Lifshitz and  L.P. Pitaevskii, {\it Physical
Kinetics}, (Pergamon Press, New York, 1981).

\bibitem{kel83}
K. F. Kelton, A. L. Greer and C. V. Thompson, J. Chem. Phys. {\bf 79}, 6261 (1983). 

\bibitem{kel91} K. F. Kelton, in {\em Solid State Physics, Vol.}
{\bf 45}, edited by H. Ehrenreich and D. Turnbull, page 75,
(Academic Press, New York, 1991).

\bibitem{cap03}
V. Capasso, ed. {\it Mathematical Modelling for Polymer Processing}.
Mathematics in Industry {\bf 2}. (Springer, Berlin 2003). 

\bibitem{gas01} U. Gasser, E.R. Weeks, A. Schofield, P.N. Pursey
and D. A. Weitz, Science {\bf 292}, 258 (2001).

\bibitem{isr91} J.N. Israelachvili, {\it Intermolecular and
Surface Forces}, 2nd. ed. (Academic Press, New York, 1991).

\bibitem{neu02}
J. C. Neu, J. A. Ca\~nizo and L. L. Bonilla, Phys. Rev. E {\bf 66}, 061406 (2002).

\bibitem{LS} I. M. Lifshitz and V. V. Slyozov, J. Phys. Chem.
Solids {\bf 19}, 35 (1961).

\bibitem{XH} S. Q. Xiao and P. Haasen, Acta metall. mater. {\bf
39}, 651 (1991).

\bibitem{mar96}  S. P. Marsh and M. E. Glicksman, Acta Mater.\
{\bf 44}, 3761 (1996).

\bibitem{pen97}
O. Penrose, J. Stat. Phys. {\bf 89}, 305 (1997).

\bibitem{juanjo}  J. J. L. Vel\'azquez, J. Stat. Phys. {\bf 92}, 195 (1998).

\bibitem{nie03}
B. Niethammer, J. Nonlin. Sci. {\bf 13} 115 (2003). 

\bibitem{pen83}   
O. Penrose and A. Buhagiar, J. Stat. Phys.
{\bf 30}, 219 (1983); O. Penrose, J. L. Lebowitz, J. Marro, M.
Kalos and J. Tobochnik , J. Stat.  Phys. {\bf 34}, 399 (1984).

\bibitem{gan99} V. Ganesan and H. Brenner, Phys. Rev. E {\bf
59}, 2126 (1999).

\bibitem{tur49}
D. Turnbull and J. C. Fisher, J. Chem. Phys. {\bf 17}, 71 (1949).

\bibitem{zel43}
Ya. B. Zeldovich, Zh. Eksp. Teor. Fiz. {\bf 12}, 525 (1942), 
Acta Physicochim. URSS {\bf 18}, 1 (1943).

\bibitem{kas69}
D. Kashchiev, Surf. Sci. {\bf 14}, 209 (1969), and {\bf 18}, 389 (1969).
 
\bibitem{wu96}
 D. T. Wu, Solid State Phys. {\bf 50}, 37 (1996).

\bibitem{tri87} 
 H. Trinkaus and M. H. Yoo, Phil. Mag. A {\bf 55}, 269 (1987).

\bibitem{shn87}
 V. A. Shneidman,  Zh. Tekh. Fiz. {\bf 57}, 131 (1987) [Sov. Phys. -- Tech. Phys. {\bf 32}, 
 76 (1987)]; Zh. Tekh. Fiz. {\bf 58}, 2202 (1988) [Sov. Phys. -- Tech. Phys. {\bf 33}, 
 1338 (1988)].
 
 \bibitem{shi90}
 G. Shi, J. H. Seinfeld and K. Okuyama, Phys. Rev. A {\bf 41}, 2101 (1990); Phys. Rev. A 
 {\bf 44}, 8443 (1991). Higher order approximation in J. J. Hoyt and G. Sundar, Scr. 
 Metall. Mater. {\bf 29}, 1535 (1993).
 
 \bibitem{shn91}
 V. A. Shneidman, Phys. Rev. A {\bf 44}, 2609 (1991).

\bibitem{dem93} 
 P. Demo and Z. Koz\'\i sek, Phys. Rev. B {\bf 48}, 3620 (1993).

\bibitem{kin02}
 J.R. King and J.A.D. Wattis, J. Phys. A {\bf 35}, 1357 (2002).
 
\bibitem{mak00} 
I. L. Maksimov, M. Sanada and K. Nishioka, J. Chem. Phys. {\bf 113}, 3323 (2000).

\bibitem{shn01}
V. A. Shneidman, J. Chem. Phys. {\bf 115}, 8141 (2001).
 
\bibitem{abr65}
 M. Abramowitz and I.A. Stegun. \textit{Handbook of Mathematical Functions}. Dover,
 1965.

\end{thebibliography}
\end{document}